\documentclass[12pt]{article}%
\usepackage{amsmath,latexsym}
\usepackage{graphicx}
\usepackage{amsmath}
\usepackage{amsfonts}
\usepackage{amssymb}%
\begin{document}

\title{{Wormholes supported by a combination of  normal
  and quintessential matter in Einstein and
  Einstein-Maxwell gravity}}
   \author{
  Peter K.F. Kuhfittig\\  \footnote{kuhfitti@msoe.edu}
 \small Department of Mathematics, Milwaukee School of
Engineering,
Milwaukee, Wisconsin 53202-3109, USA}

\date{}
 \maketitle

\begin{abstract}\noindent
It is shown in the first part of this paper that a
combined model comprising ordinary and quintessential matter
can support a traversable wormhole in Einstein-Maxwell
gravity.  Since the solution allows zero tidal forces, the
wormhole is suitable for a humanoid traveler.  The second
part of the paper shows that the electric field can be
eliminated provided that enormous tidal forces are
tolerated.  Such a wormhole would still be capable of
transmitting signals.\\

\noindent
PAC numbers: 04.20.Jb, 04.20.Gz
\end{abstract}

\section{Introduction}

Traversable wormholes, first conjectured by Morris and Thorne
\cite{MT88}, are handles or tunnels in the spacetime topology
connecting different regions of our Universe or of different
universes altogether.  Interest in traversable wormholes has
increased in recent years due to an unexpected development,
the discovery that our Universe is undergoing an accelerated
expansion \cite{aR98, sP99}.  This acceleration is due to
the presence of \emph{dark energy}, a kind of negative
pressure, implying that $\ddot{a}>0$ in the Friedmann equation
$\ddot{a}/a=-\frac{4\pi}{3}(\rho+3p)$.  In the equation of
state $p=w\rho$, the range of values $-1<w<-1/3$ results in
$\ddot{a}>0$.  This range is referred to as
\emph{quintessence dark energy}.  Smaller values of $w$ are
also of interest.  Thus $w=-1$ corresponds to Einstein's
cosmological constant \cite{mC01}.  The case $w<-1$ is
referred to as \emph{phantom energy} \cite{sS05, oZ05,
fL05a, pK06, RKSG06, KRG10}.  Here we have $\rho+p<0$, in
violation of the null energy condition.  As a result,
phantom energy could, in principle, support wormholes
and thereby cause them to occur naturally.

Sections 2-4 discuss a combined model of quintessence
matter and ordinary matter that could support a wormhole
in Einstein-Maxwell gravity, once again suggesting that
such wormholes could occur naturally.  The theoretical
construction by an advanced civilization is also an
inviting prospect since the model allows the assumption
of zero tidal forces.  Sec. 4 considers the effect of
eliminating the electric field.  A wormhole solution
can still be obtained but only by introducing a
redshift function that results in enormous radial
tidal forces, suggesting that some black holes may
actually be wormholes fitting the conditions discussed
in this paper and so may be capable of transmitting
signals, a possibility that can in principle be tested.

\section{{\textbf{  The model}} }\label{S:model}

Our starting point for a static spherically symmetric
wormhole is the line element
\begin{equation}\label{E:line1}
ds^{2}=-e^{\Phi(r)}dt^{2}+e^{\Lambda(r)}dr^{2}+r^{2}
 (d\theta^{2}+\text{sin}^{2}\theta\, d\phi^{2}),
\end{equation}
where $e^{\Lambda(r)}=1/(1-b(r)/r)$.  Here $b=b(r)$ is the
\emph{shape function} and $\Phi=\Phi(r)$ is the \emph{
redshift function}, which must be everywhere finite to prevent
an event horizon.  For the shape function, $b(r_0)=r_0$,
where $r=r_0$ is the radius of the \emph{throat} of the
wormhole.  Another requirement is the flare-out condition,
$b'(r_0)<1$ (in conjunction with $b(r)<r$), since it
indicates a violation of the weak energy condition, a
primary prerequisite for the existence of wormholes
\cite{MT88}.

In this paper the model proposed for supporting the wormhole
consists of a quintessence field and a second field with
(possibly) anisotropic pressure representing normal matter.
Here the Einstein field equations take on the following form
(assuming $c=1$):

\begin{equation}
    G_{\mu\nu}=   8 \pi G (   T_{\mu\nu}+  \tau_{\mu\nu}),
         \label{Eq3}
          \end{equation}
where $\tau_{\mu\nu}$ is the energy momentum tensor of the
quintessence-like field, which is characterized by a free
parameter $w_q$ such that $-1<w_q<-1/3$.  Following
Kiselev \cite{vK03}, the components of this tensor satisfy
the following conditions:

\begin{equation}
              \tau_t^t=    \tau_r^r = -\rho_q,
         \label{Eq3}
          \end{equation}
\begin{equation}
              \tau_\theta^\theta=    \tau_\phi^\phi = \frac{1}{2}(
              3w_q+1)\rho_q.
         \label{Eq3}
          \end{equation}
Furthermore, the most general energy momentum tensor
compatible with spherically  symmetry is
\begin{equation}
               T_\nu^\mu=  ( \rho + p_t)u^{\mu}u_{\nu}
    - p_t g^{\mu}_{\nu}+ (p_r -p_t )\xi^{\mu}\xi_{\nu}
         \label{Eq3}
          \end{equation}
with $u^{\mu}u_{\mu} = -1 $.  The Einstein-Maxwell
field equations for the above metric corresponding to
a field consisting of a combined model comprising
ordinary and quintessential matter are stated next
\cite{RKCKD11, URRRNKH}.  Here $E$ is the electric
field strength, $\sigma$ the electric charge density,
and $q$ the electric charge.
\begin{equation}\label{E:Einstein1}
  e^{-\Lambda}\left(\frac{\Lambda'}{r}
  -\frac{1}{r^2}\right)+\frac{1}{r^2}=
    8\pi G\rho+8\pi G\rho_q +E^2,
\end{equation}
\begin{equation}\label{E:Einstein2}
    e^{-\Lambda}\left(\frac{\Phi'}{r}+\frac{1}{r^2}\right)
    -\frac{1}{r^2}=8\pi Gp_r-8\pi G\rho_q-E^2,
\end{equation}
\begin{equation}\label{E:Einstein3}
   \frac{1}{2}e^{-\Lambda}\left(\frac{1}{2}(\Phi')^2
   +\Phi''-\frac{1}{2}\Lambda'\Phi'+\frac{1}{r}(\Phi'
   -\Lambda')\right)=8\pi G\left(p_t+\frac{1}{2}(3w_q+1)
   \rho_q\right)+E^2,
\end{equation}
\begin{equation}\label{E:Max4}
   (r^2E)'=4\pi r^2\sigma e^{\mu /2}.
\end{equation}
Eq. (\ref{E:Max4}) can also be expressed in the form
\begin{equation}\label{E:Max5}
    E(r)=\frac{1}{r^2}\int^r_04\pi (r')^2\sigma
    e^{\mu /2}dr'=\frac{q(r)}{r^2},
\end{equation}
where $q(r)$ is the total charge on the sphere
of radius $r$.

\section{{\textbf{Solutions}}}\label{S:solutions}

We assume that for the normal-matter field we have the
following equation of state for the radial pressure
\cite{RKR09}:
\begin{equation}\label{E:EoS1}
  p_r = m \rho, \quad -1/3<m<1.
\end{equation}
For the lateral pressure we assume the equation of state
\begin{equation}\label{E:EoS2}
p_t=n\rho, \quad -1/3<n <1.
\end{equation}
Generally, $p_r$ is not equal to $p_t$, unless, of course,
$m=n$.

Following Ref. \cite{RKR09}, the factor $\sigma
e^{\mu/2}$ is assumed to have the form $\sigma_0r^s$,
where $s$ is an arbitrary constant and $\sigma_0$ is the
charge density at $r=0$.  As a result,
\begin{equation}\label{E:E}
   E(r)=4\pi\sigma_0\frac{r^{s+1}}{s+3},
\end{equation}
\begin{equation}\label{E:E2}
   E^2(r)=16\pi^2\sigma_0^2\frac{r^{2s+2}}{(s+3)^2},
\end{equation}
and
\begin{equation}\label{E:q}
   q^2(r)=16\pi^2\sigma_0^2\frac{r^{2s+6}}{(s+3)^2}.
\end{equation}
The next step is to obtain the shape function $b(r)$
by deriving a differential equation that can be
solved for $e^{-\lambda(r)}$.  The easiest way to
accomplish this is to solve Eq. (\ref{E:Einstein1})
for $8\pi G\rho$ and substituting the resulting
expression in Eq. (\ref{E:Einstein2}), which, in
turn, is solved for $8\pi G\rho_q$.  After
substituting this result in Eq. (\ref{E:Einstein3})
and making use of Eqs. (\ref{E:EoS1}) and
(\ref{E:EoS2}), we obtain the simplified form
\begin{equation}\label{E:diffequ}
(e^{-\Lambda})^{\prime} +\frac{\alpha e^{-\Lambda}}{r} =
   \frac{\beta}{r}+rE^2\gamma,
\end{equation}
where $\alpha$, $\beta$, and $\gamma$ are dimensionless
quantities given by
\begin{equation}\label{E:1alpha}
   \alpha=\frac{-n\Phi'r/(m+1)+\frac{1}{2}(3w_q+1)
   +\frac{1}{2}(3w_q+1)\Phi'r/(m+1)+r^2\frac{1}{4}(\Phi')^2
   +\frac{1}{2}r^2\Phi''+\frac{1}{2}r\Phi'}{\frac{1}{2}
   (3w_q+1)m/(m+1)+n/(m+1)+\frac{1}{4}r\Phi'
   +\frac{1}{2}}.
\end{equation}
\begin{equation}\label{E:1beta}
  \beta=\frac{\frac{1}{2}(3w_q+1)}{\frac{1}{2}
  (3w_q+1)m/(m+1)+n/(m+1)+\frac{1}{4}r\Phi'+\frac{1}{2}}
\end{equation}
and
\begin{equation}\label{E:gamma}
  \gamma=\frac{1-\frac{1}{2}(3w_q+1)}{\frac{1}{2}
  (3w_q+1)m/(m+1)+n/(m+1)+\frac{1}{4}r\Phi'+\frac{1}{2}}.
\end{equation}

Eq. (\ref{E:diffequ}) is linear and would readily yield
an exact solution provided that $\alpha$ and $\beta$ are
constants.  This can only happen if $\Phi'=\eta/r$ for
some constant $\eta$.  In the first part of this
paper we will assume that $\eta\equiv 0$, leading to the
\emph{zero-tidal-force} solution \cite{MT88}.
Whether occurring naturally or constructed by an
advanced civilization, such a wormhole would be suitable
for humanoid travelers.

Returning to Eq. (\ref{E:diffequ}) and using Eq.
(\ref{E:E2}), the integrating factor
$e^{\alpha\,\text{ln}\, r}=r^{\alpha}$ yields the solution
\begin{equation}\label{E:solution1}
    e^{-\Lambda}=\frac{\beta}{\alpha}+
    \gamma(16\pi^2\sigma_0^2)\frac{r^{2s+4}}
    {(s+3)^2(2s+4+\alpha)}+\frac{C}{r^{\alpha}},
\end{equation}
where $C$ is an integration constant.  From
$e^{-\Lambda}=1-b(r)/r$ in Sec. \ref{S:model}, we
obtain the shape function
\begin{equation}\label{E:shape1}
   b(r)=r\left[1-\frac{\beta}{\alpha}-
   \gamma(16\pi^2\sigma_0^2)\frac{r^{2s+4}}
    {(s+3)^2(2s+4+\alpha)}-\frac{C}{r^{\alpha}}
    \right].
\end{equation}

\section{ {\textbf{Wormhole structure}} }
In Eq. (\ref{E:solution1}), $C$ is an integration
constant.  So mathematically, $e^{-\lambda}$ is a
solution for every $C$, leading to $b(r)$ in Eq.
(\ref{E:shape1}).  Physically, however, $b(r)$ is
going to satisfy the requirements of a shape
function only for a range of values of $C$.  This
problem can best be approached graphically by
assigning some typical values to the various
parameters and adjusting the value of $C$, as
exemplified by Fig. 1.  First observe that
\begin{figure}[tbp]
\begin{center}
\includegraphics[width=0.6\textwidth]{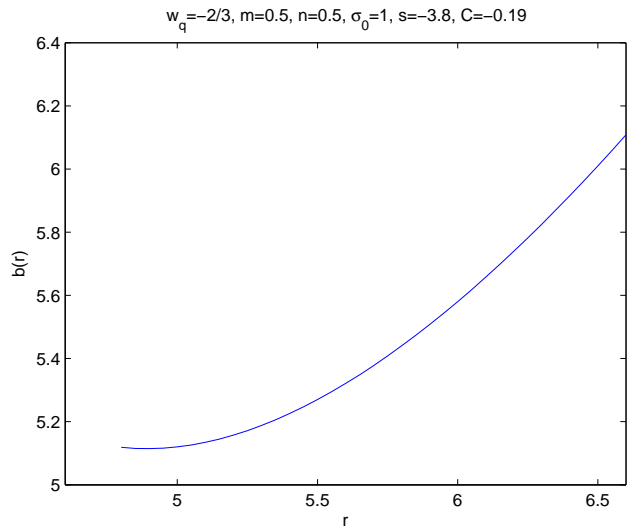}
\end{center}
\caption{The shape function.}
\end{figure}
if $\eta\equiv 0$, then $\alpha=\beta$.  For the
given values $w_q=-2/3$, $m=0.5$, $n=0.5$, $\sigma_0
=1$, and $s=-3.8$, a suitable value for $C$
is $-0.19$, as we will see.  Substituting in Eq.
(\ref{E:shape1}), we obtain
\begin{equation}\label{E:shape2}
   b(r)=127.62r^{-2.6}+0.19r^{1.75}.
\end{equation}
To locate the throat $r=r_0$ of the wormhole,
we define the function $B(r)=b(r)-r$ and determine
where $B(r)$ intersects the $r$-axis, as shown in
Fig. 2.  Observe that Fig. 2 indicates that for
\begin{figure}[tbp]
\begin{center}
\includegraphics[width=0.6\textwidth]{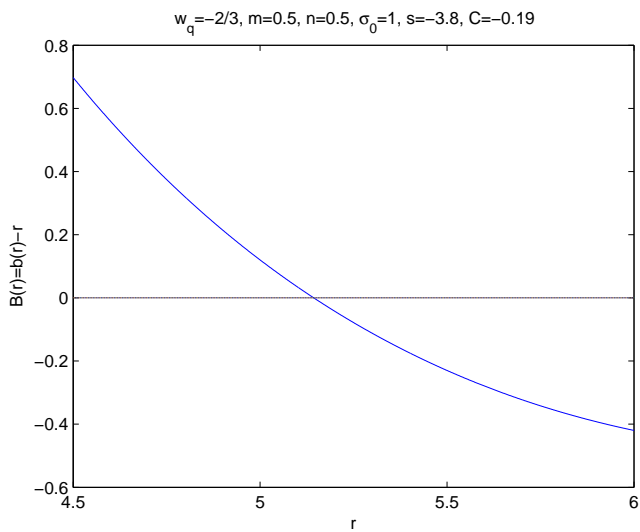}
\end{center}
\caption{$B(r)=b(r)-r$ intersects the $r$-axis at $r=r_0$.}
\end{figure}
$r>r_0$, $B(r)<0$, so that $b(r)<r$ for $r>r_0$,
an essential requirement for a shape function.
Furthermore, $B(r)$ is a decreasing function near
$r=r_0$; so $B'(r)<0$, which implies that
$b'(r_0)<1$, the flare-out condition.  With  the
flare-out condition now satisfied, the shape
function has produced the desired wormhole
structure.  For completeness let us note that
$r_0=5.143$ and $b'(r_0)=0.223$.  (Suitable
choices for $C$ corresponding to other parameters
will be discussed at the end of the section.)

To the right of $r=r_0$, $b(r)$ keeps rising,
but at $r=6.6$, $b'(r)$ is still less than
unity.  So at $r_1=6.6$, the interior shape
function, Eq. (\ref{E:shape2}), can be joined
smoothly to the exterior function
\[
   b_{\text{ext}}(r)=5.123\sqrt{r}-7.054.
\]
To check this statement, observe that
\[
   b_{\text{int}}(6.6)=b_{\text{ext}}(6.6)=6.107,
\]
while
\[
   b'_{\text{int}}(6.6)=b'_{\text{ext}}(6.6)
     =0.997.
\]
To the right of $r=r_1$, $b(r)/r\rightarrow
0$ as $r\rightarrow \infty$, so that in
conjunction with the constant redshift function,
the wormhole spacetime is asymptotically flat.
(The components $g_{\hat{\theta}\hat{\theta}}$
and $g_{\hat{\phi}\hat{\phi}}$ are already
continuous for the exterior and interior
components, respectively
\cite{LLQ03, LL04, pK09}.)

Returning to Eq. (\ref{E:shape1}), an example
of an anisotropic case is $m=0.6$, $n=0.3$,
$w_q=-2/3$, $\sigma_0=1$, and $s=-3.8$; a
suitable choice for $C$ is $-0.12$.  The
result is
\[
   b(r)=160.92r^{-2.6}+0.12r^2.
\]
Here $r_0=5.58$ and $b'(r_0)=0.48$.

An example of a value of $w_q$ closer to $-1$,
the lower end of the quintessence range, is the
following: $w_q=-0.8$, $m=n=0.5$, $\sigma_0=1$,
and $s=-3.5$.  Letting $C=-0.04$, the shape
function is
\[
   b(r)=429.52r^{-2}+0.04r^{13/6}.
\]
This time $r_0=10.32$ and $b'(r_0)=0.53$.
\\
\\
\textbf{Summary:} The emphasis in this paper is 
on the isotropic case $m=n$ since a cosmological 
setting assumes a homogeneous distribution of 
matter.  For $w_q=-1$, which is
considered to be the best model for dark energy
\cite{rB08}, $\alpha$, $\beta$, and $\gamma$,
and hence $b=b(r)$, are all independent of
$m$ and $n$.  This independence can yield a
valid solution to the field equations that
is consistent with Eqs. (\ref{E:EoS1}) and
(\ref{E:EoS2}) describing ordinary matter.
In particular, if $p=p_r=p_t$, then
$\rho+p>0$ under fairly general conditions.

\section{Could the electric field be eliminated?}

The purpose of this section is to study
conditions under which a combined model of
quintessential and ordinary matter may be
sufficient without the electric field $E$.

If $E$ is eliminated, then the assumption of
zero tidal forces becomes too restrictive.  So
we assume that $\Phi'=\eta/r$ for some nonzero
constant $\eta$.  This, in turn, means that
\begin{equation}\label{E:redshift}
  e^{\Phi}=A_1r^{\eta}.
\end{equation}
Now Eq. (\ref{E:diffequ}) yields
\begin{equation}
   e^{-\Lambda}=\frac{\beta}{\alpha}
   +\frac{A_2}{r^{\alpha}}.
\end{equation}
Both $A_1$ and $A_2$ are positive integration
constants.  (The reason that $A_2$ has to be
positive is that $\beta$ is close to zero
whenever $w_q$ is close to $-1/3$); $\alpha$
and $\beta$ now become (for $\eta\neq 0$)
\begin{equation}\label{E:2alpha}
\alpha = \frac{\frac{1}{2}(3w_q+1) +
  \frac{\eta^2}{4}+\eta[\frac{1}{2}(3w_q+1)
  -n]/(m+1)}
   {\frac{\eta}{4}+ \frac{1}{2}+
[\frac{1}{2}(3w_q+1)m+n]/(m+1)}
\end{equation}
and
\begin{equation}\label{E:2beta}
\beta = \frac{\frac{1}{2}(3w_q+1)}
  {\frac{\eta}{4}+ \frac{1}{2}+
[\frac{1}{2}(3w_q+1)m+n]/(m+1)}.
\end{equation}
The last two equations are similar to those in
Ref. \cite{RKCKD11}, which deals with galactic
rotation curves.

As noted in Sec. \ref{S:model}, the shape function $b=b(r)$
is obtained from $e^{-\Lambda(r)}$, so that
\begin{equation}\label{E:bprime}
  b(r)=r(1-e^{-\Lambda(r)})=r\left(1-\frac{\beta}{\alpha}
  -\frac{A_2}{r^{\alpha}}\right).
\end{equation}
To meet the condition $b(r_0)=r_0$, we must have
\[
   1-\frac{\beta}{\alpha}-\frac{A_2}{r_0^{\alpha}}=1.
\]
Solving for $r_0$, we obtain the radius of the throat:
\begin{equation}\label{E:rzero}
   r_0=\left(-\frac{\alpha}{\beta}A_2\right)^{1/\alpha}.
\end{equation}
Since $A_2>0$, $\alpha$ and $\beta$ must have opposite
signs.  From
$b(r)=r(1-\beta/\alpha -A_2/r^{\alpha})$, we have
\begin{equation*}
  b'(r_0)=1-\frac{\beta}{\alpha}-A_2(1-\alpha)
       r_0^{-\alpha}
\end{equation*}
and, after substituting Eq. (\ref{E:rzero}),
\[
  b'(r_0)=1-\frac{\beta}{\alpha}-A_2(1-\alpha)
  \left(-\frac{\beta}{\alpha A_2}\right),
\]
which simplifies to $b'(r_0)=1-\beta$.  It
follows immediately that if $\beta<0$, then
$b'(r_0)>1$, so that the flare-out condition
cannot be met.  To get a value for $\beta$
between 0 and 1, the exponent $\eta$ in the
redshift function, Eq. (\ref{E:redshift}),
has to be negative and sufficiently large in
absolute value.  Such a value will cause
$\alpha$ to be negative, which can best be
seen from a simple numerical example: for
convenience, let us choose $w_q=-1$, the
lower end of the quintessence range, and
$m=n=0.1$.  Then we must have $\eta<-6$.
The result is a large positive numerator
in Eq. (\ref{E:2alpha}) because the last
term is positive and $\eta^2/4$ is large.
So $\alpha$ and $\beta$ have opposite
signs, as expected.  (Observe that for the
isotropic case, if $w_q=-1$, then the
values of $\alpha$ and $\beta$ are
independent of $m$ and $n$.)

Continuing the numerical example, if we
let $A_2=1$ and $\eta=-7$, then $\beta=0.8$,
$\alpha= -14.6$, and
\[
    b(r)=1.055r-r^{15.6}.
\]
From Eq. (\ref{E:rzero}), $r_0= 0.820$,
while $b'(r_0)= 1-\beta=0.2$.  As we have
seen, $b'(r_0)$ is independent of $A_2$.
So we are free to choose a smaller value
in Eq. (\ref{E:rzero}) to obtain a larger
throat size.

We conclude that we can readily find an
interior wormhole solution around $r=r_0$
without $E$, provided that we are willing
to choose a sufficiently large (and
negative) value for $\eta$, resulting in
what may be called an unpalatable shape
function:
$\Phi= \text{ln}\,A_1+\eta\,\text{ln}\,r$.
At the throat, $|\Phi'|=|\eta/r_0|$, which
indicates the presence of an enormous
radial tidal force, even for large throat
sizes.  (Recall that from Ref. \cite{MT88},
to meet the tidal constraint, we must have
roughly $|\Phi'|<(10^8\,\text{m})^{-2}$.)
Such a wormhole would not be suitable for
a humanoid traveler, but it may still be
useful for sending probes or for
transmitting signals.

The enormous tidal force is actually
comparable to that of a solar-mass black
hole of radius 2.9 km near the event horizon,
making the solution physically plausible:
since we have complete control over $b'(r_0)$
and $r_0$, we are not only able to satisfy the
flare-out condition but we can place the throat
wherever we wish.  Moreover, the assumption
$w_q=-1$ is equivalent to Einstein's
cosmological constant, the best model for
dark energy \cite{rB08}. As noted at the 
end of Sec. 4, also physically
desirable in a cosmological setting is the
assumption of isotropic pressure, i.e.,
$m=n$ in the respective
equations of state.  As we have seen, in
the isotropic case our conclusions are
independent of $m$ and $n$.  So by
placing the throat just outside the event
horizon of a suitable black hole, it is
possible in principle to construct a
``transmission station" for transmitting
signals to a distant advanced civilization
and, conversely, receiving them.  If such a
wormhole were to exist, it would be
indistinguishable from a black hole at a
distance.  This suggests a possibility in
the opposite direction: a black hole could
conceivably be a wormhole fitting our
description.  The easiest way to test this
hypothesis is to listen for signals,
artificial or natural, emanating from a
(presumptive) black hole.


\section{Conclusion}

This paper discusses a class of wormholes supported
by a combined model consisting of quintessential
matter and ordinary matter, first in
Einstein-Maxwell gravity and then in Einstein
gravity, that is, in the absence of an electric
field.  To obtain an exact solution, it was necessary
to assume that the redshift function has the form
$e^{\Phi(r)}=A_1r^{\eta}$ for some constant $\eta$.
In the Einstein-Maxwell case, this constant could
be taken as zero, thereby producing a
zero-tidal-force solution, which, in turn, would
make the wormhole traversable for humanoid travelers.
Without the electric field $E$, the exponent $\eta$
has to be nonzero and leads to a less desirable
solution with large tidal forces.  Concerning the
exact solution, it is shown in Ref. \cite{pK11}
that the existence of an exact solution implies
the existence of a large set of additional
solutions, suggesting that wormholes of the
type discussed in this paper could occur
naturally.

It is argued briefly in the Einstein case with
a quintessential-dark-energy background that
some black holes may actually be wormholes with
enormous tidal forces, a hypothesis that may
be testable.

\end{document}